\documentclass{article}
\usepackage{amsmath,graphicx,mlspconf}
\usepackage{comment}
\usepackage{cite}
\usepackage{amssymb,amsfonts}
\usepackage{multirow}
\usepackage{subcaption}
\usepackage{booktabs}
\usepackage{bbding}
\usepackage[colorlinks,citecolor=green]{hyperref}
\usepackage{hhline,rotating}
\usepackage{pifont}

\toappear{2023 IEEE International Workshop on Machine Learning for Signal Processing, Sept.\ 17--20, 2023, Rome, Italy}

\def\x{{\mathbf x}}

\title{Lightweight Image Inpainting by Stripe Window Transformer with Joint Attention to CNN\\}
\name{Tsung-Jung Liu$^{\star}$ \qquad Bo-Wei Chen$^{\star}$ \qquad Kuan-Hsien Liu$^{\dagger}$}
\address{
\ninept $^{\star}$Department of Electrical Engineering and Graduate Institute of Communication Engineering, National Chung Hsing University, Taiwan
\\
\ninept $^{\dagger}$Department of Computer Science and Information Engineering, National Taichung University of 
}

\begin{document}\sloppy
\topmargin=0mm

\maketitle

\begin{abstract}
Image inpainting is an important task in computer vision. As admirable methods are presented, the inpainted image is getting closer to reality. However, the result is still not good enough in the reconstructed texture and structure based on human vision. Although recent advances in computer hardware have enabled the development of larger and more complex models, there is still a need for lightweight models that can be used by individuals and small-sized institutions. Therefore, we propose a lightweight model that combines a specialized transformer with a traditional convolutional neural network (CNN). Furthermore, we have noticed most researchers only consider three primary colors (RGB) in inpainted images, but we think this is not enough. So we propose a new loss function to intensify color details. Extensive experiments on commonly seen datasets (Places2 and CelebA) validate the efficacy of our proposed model compared with other state-of-the-art methods.
\end{abstract}
\section{Introduction} \label{sec1}
Image inpainting has been studied by many researchers for several years. The main goal of image inpainting is to fill up the realistic pixels in the missing region of the image and this can be applied to object removal and photo restoration. To achieve realistic results, we need to consider the following two important points: 1) the continuity of adjacent textures; 2) visually reasonable structure. All the proposed methods target at the above two points to solve the problem, such as the traditional diffusion method, patch matching method and current methods (CNN and GAN). However, they still face some limitations because convolution-based CNN has a narrow receptive field and hence it cannot get global information for the whole image. Without global information of the whole image, it is hard to repair the key edge and lines within the scene. To address this, some researchers proposed methods that utilize auxiliary information for structure recovery, e.g., edge connect (EC) \cite{nazeri2019edgeconnect}. On the other hand, some researchers proposed an attention mechanism-based model using attention scores compared with each patch to obtain global information. Suvorov \textit{et al.} \cite{suvorov2022resolution} utilized the Fast Fourier Convolution (FFC) to encode features in the frequency domain with global receptive fields for resolution-robust inpainting. Although these methods have improved the overall repair results, they require a huge computational cost. Furthermore, in recent years, transformers have also been used in the inpainting field due to their wider receptive fields than CNNs and better inpainting at low resolutions. However, transformers require a significant amount of computer memory. Therefore, it inspired us to design a lightweight transformer block with stable repair effects.

Specifically, we referred to the CSWin transformer \cite{dong2022cswin} which used stripe window self-attention to replace traditional full self-attention. The stripe window self-attention mechanism computes self-attention parallel to horizontal and vertical stripe cross-windows. Each stripe is obtained by dividing the input feature into constant-width stripes. In this way, we can achieve global attention with limited computational cost. Then we redesigned the transformer block to enhance its repair performance.

The consistency of color is another important factor to judge the quality of an image. It is easy to discern the difference between inpainted image and original image by the human eye even if there is only a small deviation in the color. While most existing methods only deal with the basic primary colors, we believe that this is not enough. If we can quickly improve color consistency in the early stage of training, the repair performance can be improved. Therefore, we transform the inpainted image to the HSV color space and compare it with the input image. In follow-up experiments, our method is confirmed to be effective.

The rest of the paper is organized as follows. In Section \ref{sec2}, we introduce the previous and state-of-the-art inpainting methods. Then we present our proposed method and loss function in Section \ref{sec3}. In Section \ref{sec4}, we exhibit our training details, experiment results, inpainting images, and ablation studies. At last, the conclusions are drawn in Section \ref{sec5}. The major contributions of this work are as follows:
\begin{itemize}
\item We propose a stripe window self-attention transformer with an efficient local enhancement position encoding. Then we redesign the transformer block to make the result better than the original method.
\item We suggest joint attention from global layers to local layers, connecting the two layers to enhance the overall consistency of repair results.
\item We propose a new HSV loss focused on color consistency in the early stage.
\item In the common dataset including Places2 and CelebA, we conduct extensive experiments to confirm that our proposed model is better than other advanced methods.
\end{itemize}
\section{Related Work}  \label{sec2}
%

 \noindent
{\bf Deep learning based inpainting.}\indent With the increasing availability of advanced hardware technology, CNN-based deep learning models have emerged as the predominant approach for image inpainting. Several deep models have been proposed in this field, including Shift-Net \cite{yan2018shift} proposed by Yan {\it et al.} and more recent models that leverage additional information such as edge information. For example, Nazeri {\it et al.} proposed Edgeconnect \cite{nazeri2019edgeconnect} and Yu {\it et al.} proposed DeepFill-V2 \cite{yu2019free} which used Canny edge detection to generate edge images. Zeng {\it et al.} \cite{zeng2021cr} proposed CRFill, which utilized auxiliary contextual reconstruction loss to encourage the generator network to borrow appropriate known regions as references for filling in a missing region. While these methods have been shown to be effective in inpainting images with complex structures such as buildings and interior spaces, they require additional stages or parameters during training. In our proposed method, we also utilize edge information, but we avoid the need for additional parameters in the model.

On the other hand, some researchers have utilized self-attention mechanisms to improve texture inpainting, such as CA proposed by Yu {\it et al.} \cite{yu2018generative} and HiFill proposed by Yi {\it et al.} \cite{yi2020contextual}. These methods compute complex attention scores to identify the most similar texture to be used in filling the missing region, and generally outperform other methods in terms of texture quality. In our proposed method, we have redesigned the attention module and incorporated wide attention to the local receptive field to enable attention sharing.

\noindent
{\bf Vision transformer.}\indent In recent years, the use of transformer models in computer vision has gained popularity. He {\it et al.} proposed the Vision Transformer (ViT) \cite{he2022masked}, which made the transformer architecture applicable to computer vision tasks. Since then, more novel transformers have been introduced, such as Dong {\it et al.}'s CSWin transformer \cite{dong2022cswin}, and some have been applied in image inpainting, such as Zheng {\it et al.}'s TFill \cite{zheng2022bridging}. Transformers are able to inpaint plausible textures for large missing regions by using their special attention mechanism. However, they require more computing resources than traditional convolutional neural networks due to their wider receptive field. In our proposed method, we have redesigned the basic transformer architecture and utilized a stripe window to divide the feature map, reducing the amount of computations and achieving better repair effects.

To summarize, this paper proposed a novel stripe window-based transformer framework for image inpainting, and enhanced it with joint attention local CNN layers. Our model focuses on the global Stripe Window Multi-Head (SWMH) transformer and CNN-based local layer. We process the global and local layer in parallel and then share the same attention information between them. In the end, we use four simple up-samples to obtain the final inpainting result. 
\begin{figure*}[htbp]
\center{\includegraphics[width=17.7cm]{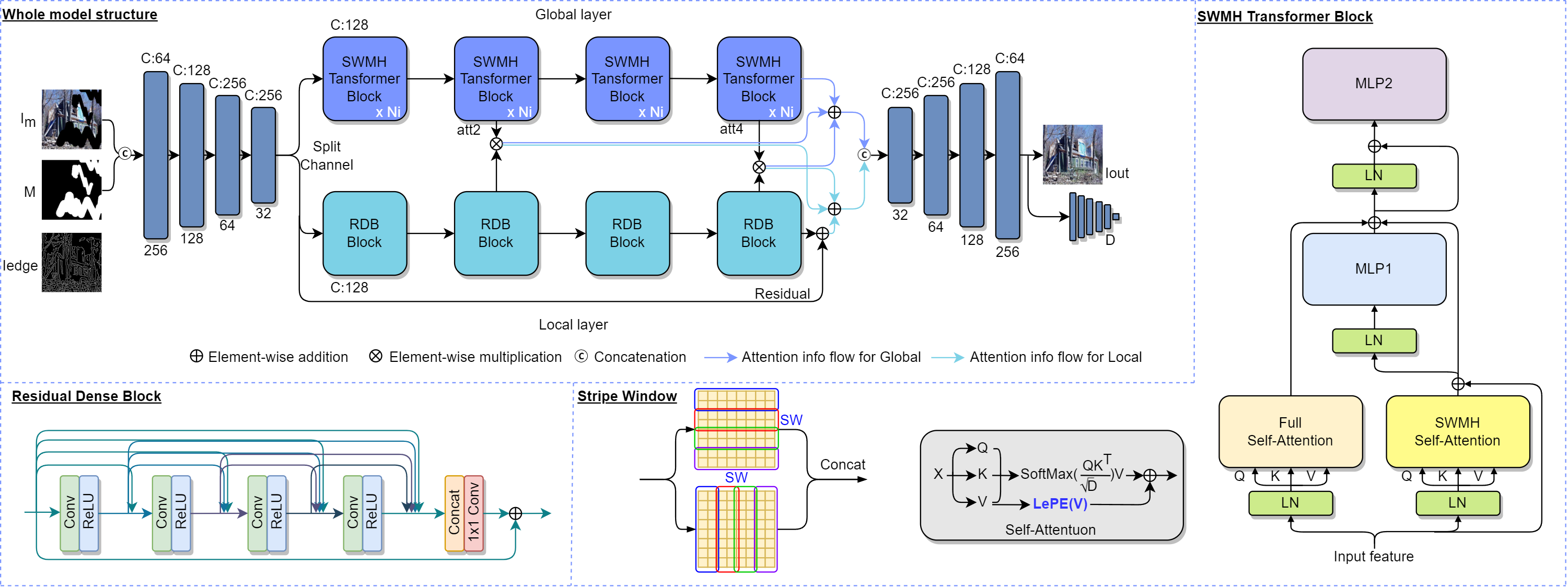}}
\caption{The overview of our proposed model. The {\bf whole model structure} shows the framework of our proposed model and the details of the joint attention between Global layer and Local layer. The input images only include $I_m$ and $M$. The $I_{edge}$ will be used in the loss function and generated by Canny \cite{ding2001canny} before training. Moreover, the right side shows the {\bf SWMH Transformer Block}. D is the normalization factor before softmax, which makes the similarity between pixels become more stable. At last, the {\bf Residual Dense Block} in the local layer is shown at the bottom left corner of the whole model.}
\label{F2}
\end{figure*}
\section{Methodology} \label{sec3}
 \noindent
{\bf Overview.}\indent The whole model of our proposed approach is shown in Fig. \ref{F2}. Given a masked image {$\bf I_m$} and a binary mask {\bf M} which are both in 256$\times$256, we concatenate them, and pass them through three downsampling CNN layers. After we downsample input image, we split the channel to global layer (i.e., SWMH transformer) and local residual in residual dense block (RRDB) \cite{wang2018esrgan} layer, where we use joint attention between the global and local layers. Each RDB block in RRDB has four consecutive Conv-ReLU. At last, we concatenate the features from both channels and then go through three upsample layers to get the inpainted image ${\bf I_{out}}$.
\subsection{{\bf Stripe Window Multi-Head (SWMH) Transformer}}\label{A}
The overall global layer of SWMH transformer is shown in Fig. \ref{F2}. The input of the global layer is a feature map with size of {\bf H$\times$W$\times$C}, where {\bf H} and {\bf W} are 32 after downsampling and the channel is 128 after the split. There are four SWMH transformer blocks in our global layer. Each block has its own multi-head and stripe window ($sw$) to reduce the amount of calculation. We set multi-head to $2, 4, 8, 16$ and $sw$ to $4, 8, 16, 32$ for the four blocks by default. The first three blocks are SWMH transformer blocks that split their channel into horizontal and vertical stripes, and then split their channel with their own multi-head again. The $sw$ will split {\bf H} or {\bf W} depending on the choice of horizontal stripes or vertical stripes. In contrast to general multi-head self-attention (MHSA), our stripe window multi-head self-attention (SWMH-SA) combines multi-head and $sw$ to greatly reduce computational complexity and achieve better inpainting effects. The last block of the SWMH transformer uses full attention because the $sw$ in the fourth block is $32$, which means the stripe window covers the whole image.\\
\noindent
{\bf SWMH Transformer Block.}\indent The structure of SWMH Transformer Block is also shown in Fig. \ref{F2}. We redesign the self-attention wiring, moving it from the first feed-forward to the beginning because we hope our self-attention block will not be influenced by the SWMH-SA. SWMH Self-Attention and Full Self-Attention will be trained from different receptive fields and then connected together with the residual link. We also add locally-enhanced positional encoding (LePE) in the transformer block to augment the positional encoding and refer to \cite{dong2022cswin} to add the LePE at the end of the transformer block but not the middle, shown on the right side of Fig. \ref{F2}. We found that self-attention needs to be calculated multiple times to get better attention information. We set the ${\bf N_{i}}$ to denote the number of repetitions.
\subsection{{\bf Joint attention}} \label{B}
We concatenate global and local layers to jointly focus on the information with different receptive fields. We expect our inpainting results to be the admixture of different receptive fields, not only single receptive field. So we collect attention from the second and fourth SWMH transformer blocks and multiply it by the corresponding RDB blocks. At last the two mixed receptive fields are added to the respective last block of the global and local layers to achieve joint attention.
\subsection{{\bf Loss Function}}
Most loss functions we adopt in this paper are the same as \cite{dong2022incremental,liu2018image,nazeri2019edgeconnect}. And we also use other losses including perceptual loss, Edge loss and HS loss which we proposed in this work. First, the ${\bf I_{out}}$, ${\bf I_{GT}}$ indicate predicted images and the ground truth, respectively. We enhance the structure of the inpainting image by using Edge loss which is $L_{edge} = \frac{1}{n} \sum_{i=1}^{n}||(I_{out}\odot M_{edge} - I_{GT}\odot M_{edge})||_{2}^2$, where $n$ represents the number of pixels in the image, and ${\bf M_{edge}} = (1 - I_{edge}) + 10*I_{edge}$, which can be seen as an edge mask to accentuate the edge structure. The ${\bf I_{edge}}$ is the image obtained from Canny edge detection \cite{ding2001canny}.

In order to improve the quality of the inpainting model, we use Perceptual loss to measure the similarity between images. We also use the mask on feature map to let Perceptual loss only focus on visible regions. The VGG-19 based perceptual loss would force the model to generate images semantically closer to the ground truth, but we notice our inpainting results have checkerboard artifacts. According to \cite{liu2018image}, checkerboard artifacts are usually caused by deconvolution and using Style loss can remove this artifact. Therefore, we use the same Style loss as \cite{liu2018image} in our total loss.

Besides focusing on texture and structure, we believe that color is as important as both. So we proposed the HS loss to measure the similarity between colors, which can be formulated as follows:
\begin{equation}
\footnotesize 
\begin{split}
&L_{HS} = \lambda_{HS}* \frac{1}{n} \sum_{i=1}^{n}||(HS_{out} - HS_{GT})||_{2}^2,\\ 
&L_{HS\_edge} = \frac{1}{n} \sum_{i=1}^{n}||(HS_{out}\odot M_{edge} - HS_{GT}\odot M_{edge})||_{2}^2,\\
&L_{HS\_T} = \lambda_{HS} * L_{HS} + \lambda_{HS\_edge} * L_{HS\_edge},
\end{split}
\end{equation}
where $\lambda_{HS} = 10$ and $\lambda_{HS\_edge} = 100$ by default. Here, {\bf HS} means $Hue, Saturation$ in HSV color space but we do not use $Value$ in the HS loss because brightness (intensity) can easily be included by other losses. If we still use the $Value$ in HS loss it will even affect our inpainting results. The $L_{HS\_edge}$ uses the edge mask, and we set it to have larger weight to enhance the boundary. We will demonstrate this in ablation experiments.

The adversarial loss includes the discriminator loss $L_D$ and the generator loss $L_G$. The adversarial loss can be indicated as
\begin{equation}
\footnotesize
\begin{split}
&L_{D} = -\mathbb{E}_{I_{GT}}[logD(I_{GT})] - \mathbb{E}_{I_{outM}}[logD(I_{out}) \odot(1 - M)] \\
&-\mathbb{E}_{I_{outM}}[log(1- D(I_{out})) \odot M], \\
&L_{G} = -\mathbb{E}_{I_{out}}[logD(I_{out})], L_{adv} = L_D +L_G+\lambda_{GP}L_{GP},
\end{split}
\end{equation}
where the PatchGAN \cite{isola2017image} based discriminator is written as {\bf D} and our proposed model can be seen as the generator {\bf G}. The $L_{GP} =  \mathbb{E}_{I_{GT}}|| \bigtriangledown_{I_{GT}}D(I_{GT}) ||^2$ is the gradient penalty and $\lambda_{GP} = 1e - 3$. We include all losses above as the total loss ${\bf L_{total}}$: 
\begin{equation}
\begin{split}
L_{total} = \lambda_{L1}&L_1 + \lambda_{edge}L_{edge} +\lambda_{perc}L_{perc}\\
&+ \lambda_{style}L_{style} + L_{HS\_T} + \lambda_{adv}L_{adv},
\end{split}
\end{equation}
where $ \lambda_{L1} = 10$, $\lambda_{edge} = 10$, $\lambda_{perc} = 0.1$, $\lambda_{style} = 250$, and $\lambda_{adv} = 10$. The above loss weights are empirically set by experiments.
\begin{table*}[ht]
\centering
\caption{Quantitative evaluation of inpainting on Places2 and CelebA datasets. We report \emph{Peak signal-to-noise ratio} (PSNR), \emph{structural similarity} (SSIM) and \emph{Learned Perceptual Image Patch Similarity} (LPIPS) metrics. The $\color{magenta}\blacktriangle$ denotes larger, and $\color{green}\blacktriangledown$ denotes lesser of the parameters compared to our proposed model. (\textbf{Bold} means the 1st best; \underline{Underline} means the 2nd best)}
\label{table1}
\scalebox{0.45}{
\begin{tabular}{rrrccccccccccccccc} 
\toprule
\multicolumn{18}{c}{\textbf{Places2~\textbar{}~CelebA}}                                                                                                                                                                                                                                                                                                                                                                                                                                                                                                                                                                                                                                                                                                                                                                                                                                                                                                                                                                                                                           \\ 
\toprule
\multicolumn{1}{l}{} & \multicolumn{1}{l}{}                                                                                                           & \multicolumn{1}{l}{}                                   & \multicolumn{1}{l}{}                                     & \multicolumn{1}{l}{}                                      & \multicolumn{2}{l}{\textbf{PSNR $\uparrow$}}                                                                          & \multicolumn{1}{l}{}                                      & \multicolumn{1}{l}{}                                      & \multicolumn{1}{l}{} & \multicolumn{1}{l}{}                                     & \multicolumn{1}{l}{}                                      & \multicolumn{2}{l}{\textbf{SSIM $\uparrow$}}                                                                          & \multicolumn{1}{l}{}                                      & \multicolumn{1}{l}{}                                      & \multicolumn{1}{l}{} & \textbf{\textbf{LPIPS$\downarrow$}}                       \\ 
\hhline{~~=======~======~=}
\multicolumn{1}{l}{} & \multicolumn{1}{c}{\begin{sideways}\begin{tabular}[c]{@{}c@{}}\\Parameters\\x10\textsuperscript{6}\end{tabular}\end{sideways}} & \multicolumn{1}{c}{\begin{sideways}mask\end{sideways}} & \begin{tabular}[c]{@{}c@{}}5\%\\$\wr$\\10\%\end{tabular} & \begin{tabular}[c]{@{}c@{}}10\%\\$\wr$\\20\%\end{tabular} & \begin{tabular}[c]{@{}c@{}}20\%\\$\wr$\\30\%\end{tabular} & \begin{tabular}[c]{@{}c@{}}30\%\\$\wr$\\40\%\end{tabular} & \begin{tabular}[c]{@{}c@{}}40\%\\$\wr$\\50\%\end{tabular} & \begin{tabular}[c]{@{}c@{}}50\%\\$\wr$\\60\%\end{tabular} &                      & \begin{tabular}[c]{@{}c@{}}5\%\\$\wr$\\10\%\end{tabular} & \begin{tabular}[c]{@{}c@{}}10\%\\$\wr$\\20\%\end{tabular} & \begin{tabular}[c]{@{}c@{}}20\%\\$\wr$\\30\%\end{tabular} & \begin{tabular}[c]{@{}c@{}}30\%\\$\wr$\\40\%\end{tabular} & \begin{tabular}[c]{@{}c@{}}40\%\\$\wr$\\50\%\end{tabular} & \begin{tabular}[c]{@{}c@{}}50\%\\$\wr$\\60\%\end{tabular} & \multicolumn{1}{l}{} & \begin{tabular}[c]{@{}c@{}}5\%\\$\wr$\\60\%\end{tabular}  \\ 
\cmidrule{1-16}\cmidrule[\heavyrulewidth]{17-18}
PM [2009]            & -~~~~                                                                                                                          &                                                        & 22.873 \textbar{} 21.440                                 & 21.523 \textbar{} 21.464                                  & 19.780 \textbar{} 20.582                                  & 17.204 \textbar{} 18.392                                  & 17.397 \textbar{} 17.531                                  & 14.921 \textbar{} 14.165                                  &                      & 0.937 \textbar{} 0.928                                   & 0.894 \textbar{} 0.909                                    & 0.882 \textbar{} 0.869                                    & 0.750 \textbar{} 0.817                                    & 0.728 \textbar{} 0.773                                    & 0.594 \textbar{} 0.661                                    &                      & ~ ~ ~-~ ~ ~\textbar{}~ ~ ~-~ ~~                           \\
CA [2018]            & 3 $\color{green}\blacktriangledown$                                                                                            &                                                        & 30.698 \textbar{} 34.559                                 & 26.575 \textbar{} 29.554                                  & 26.323 \textbar{} \textbf{29.214}                         & 22.637 \textbar{} 25.107                                  & 21.899 \textbar{} 24.317                                  & 20.366 \textbar{} 22.454                                  &                      & 0.962 \textbar{} 0.955                                   & 0.910 \textbar{} 0.928                                    & 0.903 \textbar{} \underline{0.921}                            & 0.816 \textbar{} 0.822                                    & 0.775 \textbar{} 0.811                                    & 0.710 \textbar{} 0.760                                    &                      & ~0.1831 \textbar{} 0.1226                                 \\
SN [2018]            & 55 $\color{magenta}\blacktriangle$                                                                                             &                                                        & 24.431 \textbar{} 20.753                                 & 23.057 \textbar{} 19.320                                  & 22.957 \textbar{} 18.757                                  & 22.685 \textbar{} 17.176                                  & 20.598 \textbar{} 15.718                                  & 18.306 \textbar{} 15.475                                  &                      & 0.893 \textbar{} 0.822                                   & 0.868 \textbar{} 0.818                                    & 0.842 \textbar{} 0.762                                    & 0.807 \textbar{} 0.673                                    & 0.708 \textbar{} 0.579                                    & 0.587 \textbar{} 0.537                                    &                      & 0.2221 \textbar{} 0.2647                                  \\
PC [2018]            & 49 $\color{magenta}\blacktriangle$                                                                                             &                                                        & 25.566 \textbar{} 24.902                                 & 23.429 \textbar{} 23.218                                  & 23.475 \textbar{} 23.392                                  & 24.226 \textbar{} 22.359                                  & 23.275 \textbar{} 21.005                                  & \underline{22.661 }\textbar{} 22.494                          &                      & 0.879 \textbar{} 0.859                                   & 0.845 \textbar{} 0.846                                    & 0.834 \textbar{} 0.844                                    & 0.829 \textbar{} 0.811                                    & \underline{0.803 }\textbar{} 0.765                            & 0.768 \textbar{} \textbf{0.793}                           &                      & 0.2182 \textbar{} 0.1924                                  \\
DeepFill v2 [2019]   & 4 $\color{green}\blacktriangledown$                                                                                            &                                                        & \underline{32.741 }\textbar{} \underline{33.282}                 & 28.329 \textbar{} 28.667                                  & 27.015 \textbar{} 28.634                                  & 24.117 \textbar{} 25.128                                  & 23.391 \textbar{} 24.515                                  & 21.713 \textbar{} 22.563                                  &                      & \underline{0.966 }\textbar{} \underline{0.972}                   & 0.921 \textbar{} \underline{0.924}                            & 0.904 \textbar{} 0.829                                    & 0.835 \textbar{} \underline{0.865}                            & 0.799 \textbar{} 0.815                                    & \underline{0.732 }\textbar{} 0.776                            &                      & \underline{0.1284 }\textbar{} 0.1885                          \\
HiFill [2020]        & 3 $\color{green}\blacktriangledown$                                                                                            &                                                        & 27.128 \textbar{}~ ~ ~ -~ ~ ~                            & 22.391~ \textbar{}~ ~ ~ -~ ~ ~                            & 21.906~ \textbar{}~ ~ ~ -~ ~ ~                            & 18.282 \textbar{}~ ~ ~ -~ ~ ~                             & 17.241 \textbar{}~ ~ ~ -~ ~ ~                             & 15.704~ \textbar{}~ ~ ~ -~ ~ ~                            &                      & 0.930 \textbar{}~ ~ ~-~ ~~                               & 0.825 \textbar{}~ ~ ~-~ ~~                                & 0.804~ \textbar{}~ ~ ~-~ ~~                               & 0.671 \textbar{}~ ~ ~-~ ~~                                & 0.580 \textbar{}~ ~ ~-~ ~~                                & 0.488 \textbar{}~ ~ ~-~ ~~                                &                      & 0.2506 \textbar{}~ ~ ~ -~ ~ ~                             \\
Iconv [2020]         & 30 $\color{magenta}\blacktriangle$                                                                                             &                                                        & 27.671 \textbar{} 27.174                                 & 23.629 \textbar{} 27.174                                  & 23.179 \textbar{} 26.729                                  & 20.382 \textbar{} 23.712                                  & 19.396 \textbar{} 22.841                                  & 18.313 \textbar{} 21.476                                  &                      & 0.933 \textbar{} 0.877                                   & 0.839 \textbar{} 0.877                                    & 0.822 \textbar{} 0.863                                    & 0.707 \textbar{} 0.782                                    & 0.628 \textbar{} 0.719                                    & 0.552 \textbar{} 0.666                                    &                      & 0.3810 \textbar{} 0.2517                                  \\
AOT-GAN [2020]       & 15 $\color{magenta}\blacktriangle$                                                                                             &                                                        & 31.078 \textbar{} 30.970                                 & 28.231 \textbar{} 28.558                                  & \textbf{27.947 }\textbar{} 28.389                         & \underline{24.600 }\textbar{} 25.181                          & \underline{23.741 }\textbar{} 24.539                          & 22.184 \textbar{} \underline{22.827}                          &                      & 0.950 \textbar{} 0.946                                   & 0.913 \textbar{} 0.915                                    & 0.907 \textbar{} 0.909                                    & 0.832 \textbar{} 0.854                                    & 0.791 \textbar{} 0.821                                    & 0.728 \textbar{} 0.772                                    &                      & 0.1482 \textbar{} 0.1104                                  \\
CRFill [2021]        & 4$\color{green}\blacktriangledown$                                                                                             &                                                        & \underline{32.679 }\textbar{} 32.526                         & 27.806 \textbar{} 27.443                                  & 27.339 \textbar{} 27.099                                  & 23.805 \textbar{} 23.095                                  & 22.938 \textbar{} 22.308                                  & 21.418 \textbar{} 20.522                                  &                      & 0.964 \textbar{} 0.966                                   & 0.914 \textbar{} 0.916                                    & 0.906 \textbar{} 0.908                                    & 0.828 \textbar{} 0.832                                    & 0.787 \textbar{} 0.791                                    & 0.729 \textbar{} 0.733                                    &                      & 0.1925 \textbar{} 0.1379                                  \\
TFill [2022]         & 15~$\color{magenta}\blacktriangle$                                                                                             &                                                        & \textbf{33.191 }\textbar{} \textbf{35.143}               & \underline{28.717 }\textbar{} \textbf{29.269}                 & 27.42 \textbar{} 28.664                                   & 24.43 \textbar{} \underline{25.651}                           & 23.684 \textbar{} \underline{24.517}                          & 21.915 \textbar{} 22.86                                   &                      & \textbf{0.968 }\textbar{} \textbf{0.975}                 & \underline{0.922 }\textbar{} \textbf{0.929}                   & \underline{0.911 }\textbar{} 0.915                            & \underline{0.842 }\textbar{} 0.863                            & \underline{0.803 }\textbar{} \underline{0.816}                    & 0.728 \textbar{} 0.775                                    &                      & 0.1331 \textbar{} \textbf{0.0972}                         \\ 
\cmidrule{1-16}\cmidrule[\heavyrulewidth]{17-18}
Ours [2023]          & 6~~~~                                                                                                                          &                                                        & 31.175 \textbar{} 31.782                                 & \textbf{28.718 }\textbar{} \underline{28.849}                 & \underline{27.753 }\textbar{} \underline{28.708}                  & \textbf{24.842 }\textbar{} \textbf{25.907}                & \textbf{24.127 }\textbar{} \textbf{24.616}                & \textbf{22.866 }\textbar{} \textbf{22.916}                &                      & 0.944 \textbar{} 0.947                                   & \textbf{0.923 }\textbar{} \underline{0.924}                   & \textbf{0.912 }\textbar{} \textbf{0.923}                  & \textbf{0.849 }\textbar{} \textbf{0.871}                  & \textbf{0.804 }\textbar{} \textbf{0.822}                  & \textbf{0.734 }\textbar{} \underline{0.782}                   &                      & \textbf{0.1217 }\textbar{} \underline{0.1032}                 \\
\bottomrule
\end{tabular}
}
\end{table*}
\section{Experiments}  \label{sec4}
\subsection{{\bf Datasets}}
To show the inpainting effectiveness of our proposed model, we conduct experiments on Places2 dataset. For Places2, we randomly chose 20k images from the original dataset as the training set, 5k images as the validation, and used about 4k images as the test. We use less data and the lightweight model to show our proposed approach has better robustness than other state-of-the-art huge-parameter models. For CelebA dataset, we split the dataset into 8:1:1 for training, validation and test. For all of the images in above two datasets, we only train and test them with image size 256$\times$256. For other comparison methods, we use their provided pretrained models to perform the test on the same dataset as we did.
\subsection{{\bf Reference State-of-the-Art}}
We compare the proposed model with other state-of-the-art (SOTA) methods, which include PatchMatch (PM) \cite{barnes2009patchmatch}, Contextual Attention (CA) \cite{yu2018generative}, Shift-net (SN) \cite{yan2018shift}, Partial Convolutions (PC) \cite{liu2018image}, Gated Convolution (DeepFill-v2) \cite{yu2019free}, Contextual Residual Aggregation (HiFill) \cite{yi2020contextual}, Imputed Convolution (Iconv) \cite{hukkelaas2020image}, Aggregated contextual transformations (AOT-GAN) \cite{zeng2022aggregated}, Auxiliary Contextual Reconstruction (CRFill) \cite{zeng2021cr}, Bridging Global Context Interactions (TFill) \cite{zheng2022bridging}. 
\subsection{{\bf Quantitative Comparisons}}
In Table \ref{table1}, we utilize PSNR, SSIM \cite{wang2004image} and LPIPS to assess the performance of all compared methods and our proposed approach on the two datasets with irregular masks of different masking rates. The model parameters are also shown beside each method, where the results are tested by ourselves. For two datasets, our proposed method can defeat most of compared methods in terms of these three evaluation metrics. Among them, LPIPS \cite{zhang2018unreasonable} is considered a better metric than other metrics in the inpainting field, because LPIPS used perceptual distance to compare high-level information which will be better than other low-level metrics. Hence, we also use this metric to compare the performance for all methods, which shows the robustness of our proposed model. On the other hand, our training images and steps are also less than most methods, so we can see the proposed method is effective from Table \ref{table1}.
%
\subsection{{\bf Qualitative Comparisons}}
\begin{figure*}[htbp]
\includegraphics[width=12.5cm]{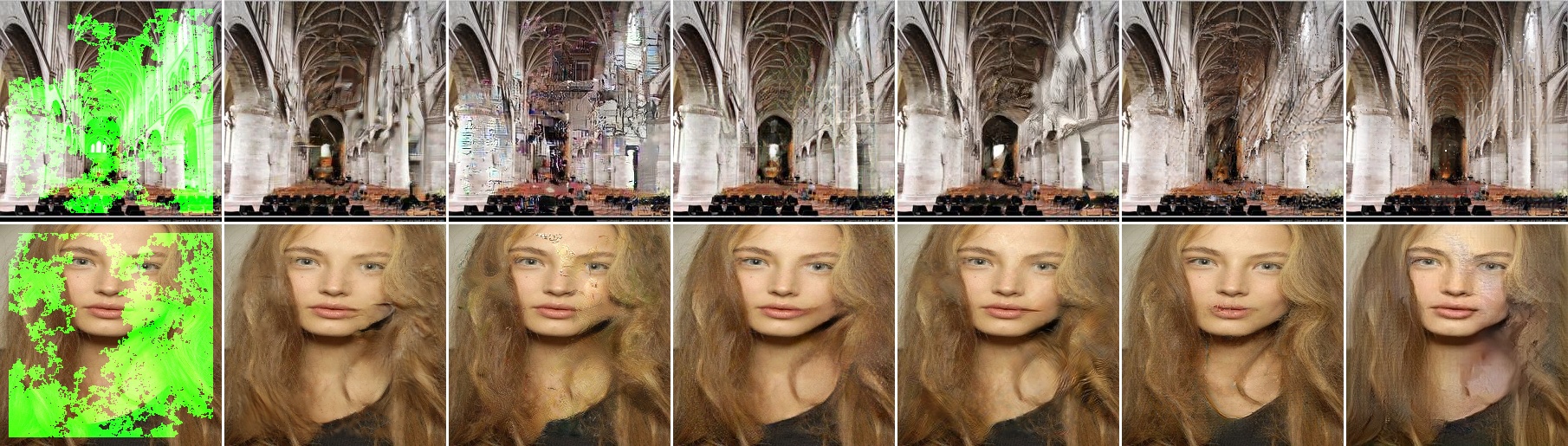}
\centering
\caption{Qualitative results of Places2 (upper half) and CelebA (lower half) datasets among SOTA methods. From left to right: Masked image, DeepFill-v2\cite{yu2019free}, Iconv\cite{hukkelaas2020image}, AOT-GAN\cite{zeng2022aggregated}, CRFill\cite{zeng2021cr}, TFill\cite{zheng2022bridging}, and Ours. Zoom-in for details.}
\label{F3}
\end{figure*}

We show the qualitative inpainting results of Places2 and CelebA in Fig. \ref{F3}. Compared with other methods, our proposed model can reconstruct similar or even more clear textures. We notice our inpainting results are slightly blurred when we focus more on the transformer and less on CNN. In the future, we will set restrictions on the local layers so that local information will not be ignored. Furthermore, our architecture is a lightweight model, which means we do not need lots of parameters, and still can achieve similar results compared to those larger models. Note that both our training data and steps are less than other methods.
\begin{table}
\centering
\caption{Ablation study of RDB, HS loss and SWMH transformer with size 256$\times$256 images on Places2 dataset.}
\label{table3}
\scalebox{0.5}{
\begin{tabular}{ccccccc|ccc} 
\toprule
w/ RDB                                             & w/ ViT                                           & w/ CSWin                                         & w/ SWMH                                          & w/ HS\_edge                                       & w/ HSV                                      & w/ HS                                                                            & \textbf{\textbf{\textbf{\textbf{PSNR$\uparrow$}}}} & \textbf{SSIM~$\uparrow$} & \textbf{\textbf{LPIPS$\downarrow$}}  \\ 
\hline\hline
x    & x                                        & x                                        & \textcolor[rgb]{0.071,0.071,0.071}{$\checkmark$} & \textcolor[rgb]{0.071,0.071,0.071}{$\checkmark$} & x                                        & \textcolor[rgb]{0.071,0.071,0.071}{}\textcolor[rgb]{0.071,0.071,0.071}{$\checkmark$} & 25.3351                                            & 0.7922                   & 0.1400                               \\
\textcolor[rgb]{0.071,0.071,0.071}{$\checkmark$}                                         & \textcolor[rgb]{0.071,0.071,0.071}{$\checkmark$} & x                                        & x                                        & \textcolor[rgb]{0.071,0.071,0.071}{$\checkmark$} & x                                        & \textcolor[rgb]{0.071,0.071,0.071}{$\checkmark$}                                     & 25.7935                                            & 0.8072                   & 0.1242                               \\
\begin{tabular}[c]{@{}c@{}}\textcolor[rgb]{0.071,0.071,0.071}{$\checkmark$}\\\end{tabular} &x                                         & \textcolor[rgb]{0.071,0.071,0.071}{$\checkmark$} & x                                        & \textcolor[rgb]{0.071,0.071,0.071}{$\checkmark$} & x                                        & \textcolor[rgb]{0.071,0.071,0.071}{$\checkmark$}                                     & 26.1027                                            & 0.8377                   & 0.1221                               \\
\textcolor[rgb]{0.071,0.071,0.071}{$\checkmark$}                                           & x                                        & x                                        & \textcolor[rgb]{0.071,0.071,0.071}{$\checkmark$} & x                                        & x                                        & x                                                                            & 26.2786                                            & 0.8459                   & 0.1212                               \\
\textcolor[rgb]{0.071,0.071,0.071}{$\checkmark$}                                           & x                                        & x                                        & \textcolor[rgb]{0.071,0.071,0.071}{$\checkmark$} & x                                        & x                                        & \textcolor[rgb]{0.071,0.071,0.071}{$\checkmark$}                                     & 26.3444                                            & 0.8422                   & 0.1193                               \\
\textcolor[rgb]{0.071,0.071,0.071}{$\checkmark$}                                           & x                                        & x                                        & \textcolor[rgb]{0.071,0.071,0.071}{$\checkmark$} & \textcolor[rgb]{0.071,0.071,0.071}{$\checkmark$} & \textcolor[rgb]{0.071,0.071,0.071}{$\checkmark$} & x                                                                            & 26.4757                                            & 0.8541                   & 0.1184                               \\ 
\toprule
\textcolor[rgb]{0.071,0.071,0.071}{$\checkmark$}                                           & x                                        & x                                        & \textcolor[rgb]{0.071,0.071,0.071}{$\checkmark$} & \textcolor[rgb]{0.071,0.071,0.071}{$\checkmark$} & x                                        & \textcolor[rgb]{0.071,0.071,0.071}{$\checkmark$}                                     & \textbf{26.5801}                                   & \textbf{0.8611}          & \textbf{ 0.1156 }                    \\
\bottomrule
\end{tabular}
}
\end{table}

\begin{figure}[htbp]
\includegraphics[width=8.5cm]{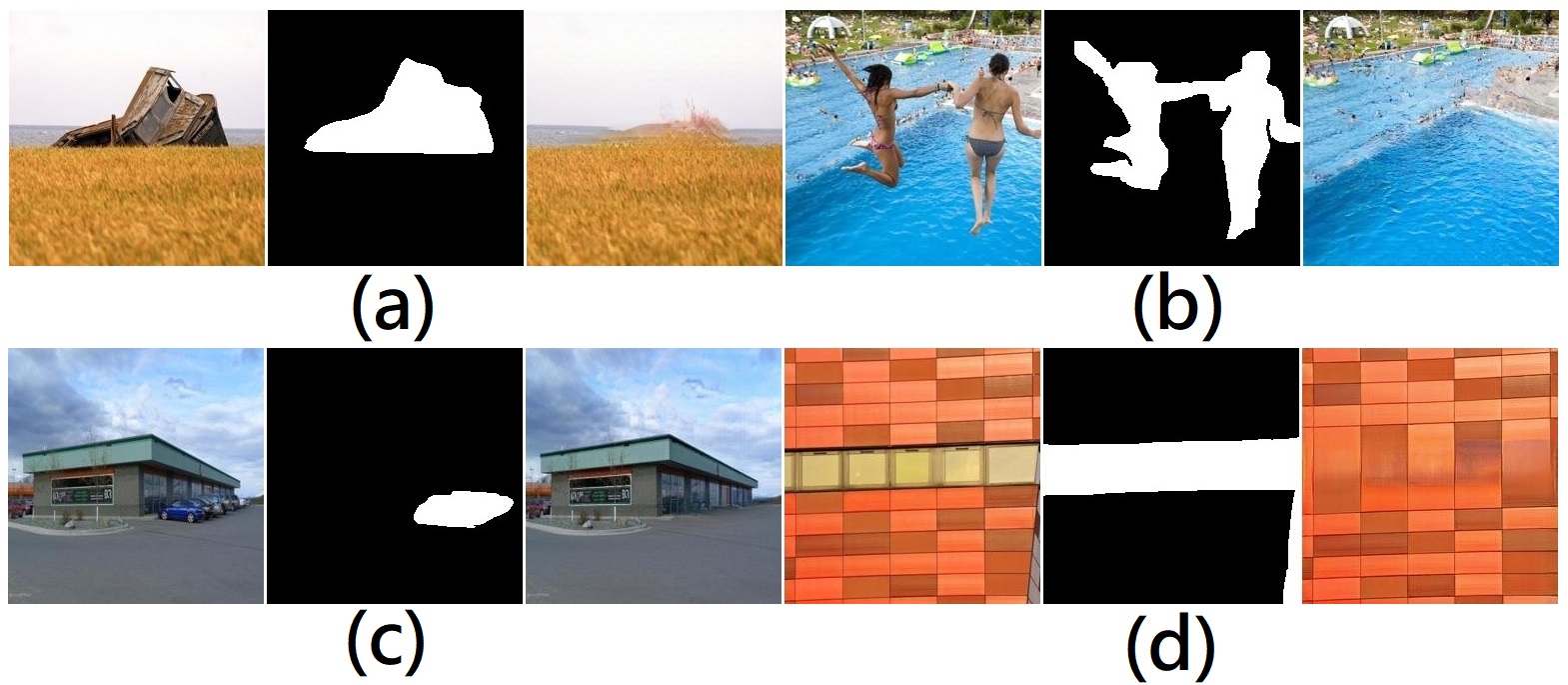}
\centering
\caption{Object removal (size 256$\times$256) results. In (a)-(d), from left to right: Ground-truth image, mask, object removal result.}
\label{F4}
\end{figure}
\begin{figure}[htbp]
\includegraphics[width=7cm]{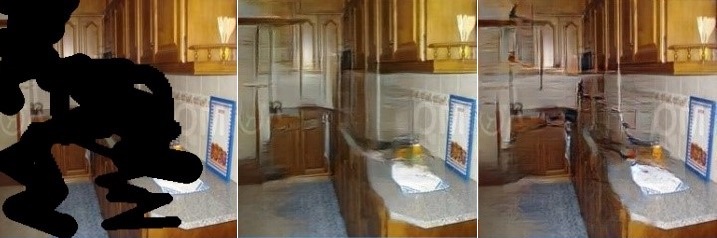}
\centering
\caption{Ablation study of color deviation on inpainted images. From left to right: Masked images, w/o $L_{HS\_T}$ loss, and w/ $L_{HS\_T}$ loss.}
\label{F5}
\end{figure}
\subsection{{\bf Ablation Study}}
To confirm our proposed module and new loss function are useful in the proposed architecture, we separately test them in the ablation experiments. We test the stability of the SWMH transformer and the redesign in Table \ref{table3}. We retrained the CSWin transformer without redesign and original transformer \cite{he2022masked} separately and compared them with our redesigned SWMH transformer. For the results shown in Table \ref{table3}, our proposed approach has the best PSNR, SSIM, and LPIPS.

We also conduct experiments for HS loss in Table \ref{table3}. We noticed the Value (V) of HSV can easily be learned in $L_1$ and other losses. If we still consider V in $L_{HS}$, it will influence the balance of the inpainting result, as shown in the table. We show the color deviation between with and without $L_{HS}$ at early training steps in Fig. \ref{F5}. We can see the color of the inpainting results in the early 50 training steps, which shows the one with $L_{HS}$ is more close to the ground truth than without $L_{HS}$, and the known region and the missing region are more consistent when using $L_{HS}$.

At last, in order to confirm the joint attention with local layer is effective, we remove the whole local layer and only keep the global layer (i.e., w/o RDB). We can see the inpainting results become worse without local layers. 
\subsection{{\bf Object removal}}
Moreover, we demonstrate our model has practical applications in Fig \ref{F4}. In this figure, we show object removal and background inpainting results.

\section{Conclusion}  \label{sec5}
\vspace{-0.5cm}
In this paper, we propose a lightweight joint attention transformer. We use transformer-based architecture to get wide receptive field information and cooperate with local layers with RRDB by joint attention with each other. Our proposed HS loss can stabilize the colors in early training steps and eventually further improve the inpainting performance. We refer to the CSWin transformer and proposed the SWMH transformer block to not confuse the two self-attentions and achieve significant improvements. Our experiments demonstrate that the proposed model using small amount of parameters can still generate similar or even better inpainting results than other SOTA methods. Those large models do have an advantage in details but not every researcher has enough hardware support. Therefore, we propose this approach to demonstrate small models are also able to compete with large models.
\bibliographystyle{IEEEbib}
\selectfont
\bibliography{mlsp_2023}
\end{document}